\documentclass[12pt]{article}
\usepackage{epsfig}

\begin{document}
\newcommand{\ice}[1]{\relax}
\newcommand{\al}{\alpha} 
\newcommand{\be}{\begin{equation}}
\newcommand{\ee}{\end{equation}}
\newcommand{\ba}{\begin{eqnarray}}
\newcommand{\ea}{\end{eqnarray}}    
\newcommand{\alsb}{\left( \frac{\alpha_s}{\pi} \right) }
\newcommand{\als}{ \frac{\alpha_s}{\pi}  }
\newcommand{\MSsch}{\overline{\rm{MS}}}
\newcommand{\mts}{M_\tau^2}
\newcommand{\nn}{\nonumber}
\newcommand{\ap}{a^\prime}
\renewcommand{\topfraction}{1.0}

\begin{center} 
{\Large \bf Determination of the strange quark 
mass from Cabibbo-suppressed tau decays 
with resummed perturbation theory 
in an effective scheme 
}
\vspace{0.6truecm}

{\large J.G.~K\"orner$^1$, F.~Krajewski$^1$, A.A.~Pivovarov$^{1,2}$}\\[.1cm]
$^1$ Institut f\"ur Physik, Johannes-Gutenberg-Universit\"at,\\[-.1truecm]
  Staudinger Weg 7, D-55099 Mainz, Germany\\[.2truecm]
$^2$ Institute for Nuclear Research of the\\[-.1truecm]
  Russian Academy of Sciences, Moscow 117312
\vspace{0.6truecm}
\end{center}

\vskip 1cm
\centerline {\bf Abstract}
We present an analysis of the $m_s^2$-corrections 
to Cabibbo-suppressed $\tau$ lepton decays employing contour 
improved resummation within an effective scheme
which is an essential new feature as compared to previous analyses.
The whole perturbative QCD dynamics of the $\tau$-system is 
described by the $\beta$-function of the effective coupling constant
and by two $\gamma$-functions for the effective mass
parameters of the strange quark in different spin channels.
We analyze the stability of our results with regard to high-order 
terms in the perturbative expansion of the renormalization group functions.
A numerical value for the 
strange quark mass in the $\MSsch$ scheme is extracted
$m_s(M_\tau)=130\pm 27_{\rm{exp}}\pm 9_{\rm{th}}~{\rm MeV}$.
After running to the scale $1~{\rm GeV}$ this translates into 
$m_s(1~{\rm GeV})=176 \pm 37_{\rm{exp}}\pm 13_{\rm{th}}~{\rm MeV}$.
\thispagestyle{empty}


\newpage

\section{Introduction}
The $\tau$-system offers the possibility to confront QCD with 
experiment in the low energy region. The high precision 
of experimental data and good accuracy of theoretical results 
make $\tau$-physics an important 
testing ground for QCD \cite{PDG,exp1,exp2}.
Theoretically the observables of the $\tau$-system are related to the
moments of the spectral density of a correlator of hadronic currents 
which can be reliably calculated within perturbation theory
\cite{cont,cont1,cont2}.
Therefore, the $\tau$-system observables were extensively
studied during the past few years
within the operator product expansion (OPE)
which is a general approach to analyzing the properties
of current correlators
\cite{SchTra84,Bra88Bra89,NarPic88,pivNew,BraNarPic92}. 
The perturbation theory (PT) series in QCD appear to be  
asymptotic and the ultimate accuracy they 
can provide depends on the concrete 
numerical value of the expansion parameter --
the strong coupling constant $\al_s(E)$ at a relevant energy $E$.
This limits the theoretical accuracy which can be obtained
within the finite order perturbation theory (FOPT) analysis. 
In the case of the $\tau$-system 
the strong coupling constant $\al_s(M_\tau)$ is not small
at the scale of the $\tau$ lepton mass $M_\tau$ 
which can lead to an asymptotic growth of terms of the perturbation 
theory series already at a rather low (third-fourth) order of 
PT expansion.
Judging from the analysis of the moments 
of the hadronic spectral density in the finite 
energy interval \mbox{$(0,M_\tau^2$ )} within finite order
perturbation theory there are strong indications 
that the ultimate theoretical 
accuracy for the $\tau$ lepton decay observables has 
already been reached at next-to-next-to-next-to-leading order 
($\rm{N}^3\rm{LO}$) or $\al_s^3$
which is the highest order of PT expansions 
presently available \cite{onetwo}.
The convergence behavior of the perturbation series for 
the $\tau$ lepton observables depends on the 
region of the spectral density which is being probed:
if the low energy region is suppressed 
(as for high moments of the spectral density)  
the asymptotic limit of the series moves to higher order terms.
The expansion of the correlator 
in $m_s^2$ makes the explicit convergence of the PT series 
for the coefficient
functions of consecutive $m_s^2$ corrections slower
\cite{cheta3}. The reason for such a behavior is quite 
obvious -- higher order
terms of $m_s^2$ expansion of the correlator are more 
sensitive to the low energy region of integration in Feynman 
diagrams that is not described by perturbation theory. 
Thus, the PT expansions for observables in the $\tau$-system
seem to be at the edge of asymptotic growth in the N$^3$LO.   
At the same time, the present accuracy of experimental data 
for some observables of the $\tau$-system
is already comparable with the ultimate 
theoretical accuracy reachable in FOPT \cite{exp1,exp2}.
This raises the problem of obtaining 
more precise theoretical formulas. Higher order terms of FOPT 
(though they are very desirable and provide 
additional information) 
will not give more precise results for PT series if the 
asymptotic limit is already reached and must simply be 
discarded in FOPT applications. To catch up with improving 
experimental accuracy 
it is necessary to find a way of interpretation of the perturbation 
theory series that allows one to reach a theoretical 
precision comparable to that of the experimental data.
One possibility to extract numerical 
results from a perturbation theory series which is explicitly divergent
at finite orders is to apply a resummation procedure 
(see e.g. \cite{Zakh}). This is more sophisticated 
than just summing the consecutive terms of the perturbative expansion
up to some finite order 
but it requires some knowledge (or assumptions) about 
the behavior of infinite number of terms of the expansion.
The choice of a resummation procedure is not 
unique and there are many ways to resum or improve the convergence of 
an asymptotic series e.g. \cite{renRS,pivrho}. 
We think that there are two important criteria for the 
choice of an appropriate resummation procedure
for perturbation theory series of physical observables: 
the renormalization group structure of the PT series
should be respected and the definition of the 
effective parameters used for the description of the 
observables should be physically motivated. 
Led by these two criteria we use 
contour improved perturbation theory (CIPT) \cite{pivNew,Pivtau}
(see also \cite{DP}) in an effective scheme (not $\MSsch$) to resum 
perturbation theory contributions in all orders. 
Within CIPT, the procedure of resummation is, however, not unique. 
A freedom in the choice of different 
renormalization schemes still remains within CIPT, which will affect 
the final numerical results \cite{groote}.
We choose an effective scheme which we consider to be
natural and the simplest one for the $\tau$-system.
Note that the choice of an adequate scheme is also 
dictated by the way the system is to be described. 
In the FOPT analysis of the moments, the contour integration is
completely (mathematically) equivalent to the integration along the
physical cut. Therefore, the basic object for FOPT analysis
is the PT spectral density of the correlator which determines 
the effective coupling in this procedure \cite{onetwo}.
In CIPT, the correlator is the basic perturbation theory 
object which is naturally normalized in Euclidean domain.
Within the effective scheme approach all PT 
corrections to the correlator are absorbed
into the definition of the effective parameters of the system
\cite{effsch,ksch,kksch,effDh,brodsky,prl}.
In the massless case the effective coupling constant 
is the only relevant parameter.
If the $m_s$-corrections to the correlator are included
then two additional parameters $m^2_q$ and $m^2_g$ related 
to mass corrections to different spin structures 
of the correlator should be considered. These parameters are 
perturbatively related to a finite strange quark mass. 
In the massless case the effective scheme resummation
analysis along these lines was done in \cite{groote}. 
In this paper the resummation analysis 
in the effective scheme is extended to $m_s^2$ corrections. 
An analysis of the $m_s^2$ 
corrections within CIPT in the $\MSsch$ scheme
was previously performed in refs.~\cite{msNPB,pichprades}.
Our results of resummation within the effective scheme approach
confirm the main conclusions of the paper \cite{msNPB}
and are consistent with the results of
ref.~\cite{pichprades}
though the direct comparison is a bit complicated because we
use a different parameterization of the relevant 
observables. We also use
different moments for the final determination of a numerical
value for $m_s$ as compared to ref.~\cite{pichprades} -- 
the detailed discussion of our choice is given in the text.  
The uncertainty of our new results is smaller than the
conservative error bars given in ref.~\cite{msNPB}
on the base of an analysis of the explicit convergence of PT
series. In order to get an understanding of the reliability of our
present procedure we evaluate the stability of our results 
with regards to higher order corrections
to the renormalization group functions. 

The paper is organized as follows.
In Sect. 2 we set up the stage for resummation in an effective
scheme. We define our basic quantities and calculate their
respective anomalous dimensions (beta-functions) that
determine the running along the integration contour. The known
results of PT calculation in the $\MSsch$ scheme 
are used but the expressions
for the corresponding renormalization group (RG) 
functions of effective parameters are
new and specific for the effective scheme defined in the paper.
In Sect. 3 we present the results of theoretical calculations
of the moments of the spectral density of $\tau$ decays in
the effective scheme. The PT expressions are given in the effective
scheme which is a new result. We introduce the effective mass parameters
for the channels with spin 1 and spin 0 and give explicit
formulas expressing these parameters through the standard $\MSsch$
mass $m_s$ in a resummed form. Some nonPT corrections are taken at the
leading order only and are therefore scheme independent:
they coincide with the results already published in the
literature (see, e.g.~\cite{pichprades}).
In Sect. 4 we extract a numerical value for the
strange quark mass from experimental data and compare our results
with conclusions of earlier analyses. 
Sect. 5 gives our conclusion. In an Appendix the RG functions of
the effective scheme are explicitly given through the coefficient
of the $\MSsch$-scheme RG functions known from the literature.

\section{Resummation in an effective scheme}
In the first step of our analysis of the $\tau$-system we define an
effective scheme in which all higher order PT corrections
to the correlator of hadronic currents 
are absorbed into the effective coupling and two 
effective mass scales,
or two coefficient functions of mass corrections.
If such a scheme is used for describing the $\tau$-system the only 
perturbative objects are one effective 
$\beta$-function for the coupling constant 
and two effective $\gamma$-functions for the coefficient functions
of mass corrections.
Using these three functions we determine the evolution of the 
effective coupling and of the two coefficient functions of mass
corrections on the contour in the complex plane of squared momentum.
Within our procedure the $\beta$- and $\gamma$-functions 
are treated as exact functions 
which is a standard understanding of renormalization group (RG)
summation. Having the explicit solutions for the running coupling 
and the mass coefficient functions in hand we determine 
observables such as the moments of the decay rate simply by 
integrating the coefficient functions with a weight function 
specific to a chosen observable. 

\subsection{Definition of an effective scheme}
The basic theoretical quantity for describing 
the $\tau$ semileptonic decays is the correlator of 
two hadronic currents
\begin{equation}
\label{corrF}
\Pi_{\mu\nu}(q) = i \int dx e^{iqx}
\langle T j_{\mu}(x) j_{\nu}^{\dagger} (0) \rangle
= \frac{N_c}{6\pi^2}( q_{\mu}q_{\nu} \Pi_q(q^2)+g_{\mu\nu}  \Pi_g(q^2) )  
\label{correlator}
\end{equation}
with $j_{\mu}(x) = \bar{u}\gamma_{\mu}(1-\gamma_5) s$.
Here $\Pi_q(q^2)$ and $\Pi_g(q^2)$ are scalar invariant functions,
$N_c=3$ is the number of colors in QCD. 
We work within QCD with three light quarks.
The correlator is normalized 
to unity in the leading (parton model) approximation
with massless quarks.
By expanding $\Pi_q(q^2)$ and $\Pi_g(q^2)$ 
in terms of powers of $m_s^2/q^2$ and keeping 
only the first order term in this expansion one has
\be
\label{expmq}
\Pi_q(q^2)=\Pi(q^2)+3\frac{m_s^2}{q^2}\Pi_{mq}(q^2)\, ,
\ee
\be
\label{expmg}
\Pi_{g}(q^2)=-q^2\Pi(q^2)+\frac{3}{2} m_s^2\Pi_{mg}(q^2)
\ee
where $\Pi(q^2)$ is an invariant function 
already known from the mass zero case. The functions 
$\Pi_i(Q^2) (i = m_q,m_g)$ with $Q^2=-q^2$ are computable in
perturbation theory in the deep Euclidean region 
$Q^2\rightarrow \infty$.  
The results of perturbation theory calculations
for the correlator given in eqs.~(\ref{expmq},\ref{expmg})
were obtained in refs.~\cite{phys_report,eek20,eek21,eek2c} 
and have already been used in the FOPT analysis \cite{onetwo}.

We define new effective quantities 
$a$, $m_q^2$, $m_g^2$ such that all information 
from perturbation theory calculations for the functions 
$\Pi(Q^2)$, $\Pi_{mq,mg}(Q^2)$ is absorbed into the evolution
of these new quantities which is determined by the effective 
$\beta$- and $\gamma$-functions.
For the mass corrections we introduce two 
different coefficient functions (we sometimes call them 
the effective mass parameters) because the correlator 
in eq.~(\ref{correlator}) 
is not transverse if corrections of the order $m_s^2$
are taken into account. Our definitions of effective quantities 
$a$, $m_q^2$, $m_g^2$ are 
\begin{eqnarray} 
\label{effdef}
-Q^2\frac{d}{dQ^2}\Pi(Q^2)
&=&1 + a(Q^2)\, ,   \nn \\
- m_s^2(\mts) Q^2\frac{d}{dQ^2}\Pi_{mg}(Q^2)
&=& m_g^2(\mts)C_g(Q^2)\, , \\
m_s^2(\mts) \Pi_{mq}(Q^2)
&=& m_q^2(\mts)C_q(Q^2) \nn \, .
\end{eqnarray}
Here $C_{q,g}(Q^2)$ are coefficient functions of mass corrections.
They are conveniently normalized by the requirement $C_{q,g}(\mts)=1$.
 
In terms of the $\MSsch$ scheme quantities 
$\al_s\equiv \al_s(\mts)$ and $m_s\equiv m_s(\mts)$
the effective parameters in eq.~(\ref{effdef}) read
\ba\label{effparam}
a(\mts)&=& 
\als+k_1\alsb^2 + k_2\alsb^3 + k_3\alsb^4 + {\cal O}(\al_s^5)  \, , \\  
\label{effmg}
m_g^2(\mts)&=& 
m_s^2(\mts)(1+\frac{5}{3}\als +k_{g1}\alsb^2 +k_{g2}\alsb^3 
+ {\cal O}(\al_s^4))\, ,\\
\label{effmq}
m_q^2(\mts)&=& 
m_s^2(\mts)(1+\frac{7}{3}\als + k_{q1}\alsb^2+ k_{q2}\alsb^3
+ {\cal O}(\al_s^4))\, .
\ea
Numerical values for the coefficients $k_3$, $k_{q2}$
are unknown though their estimates within various approaches
can be found in the literature.
We have written explicitly first coefficients for mass corrections, 
$k_{g0}=5/3$, $k_{q0}=7/3$.
Further references concerning numerical values for 
the known coefficients $k_{1,2}$, $k_{g1,g2}$,  
$k_{q1}$ are given in Appendix.

\subsection{Running of the effective coupling $a(Q^2)$ and \\
the mass coefficient functions $C_{q,g}(Q^2)$}
The behavior of the effective coupling $a(Q^2)$ and 
the coefficient functions of the mass parameters $C_{q,g}(Q^2)$
is determined by the effective beta and gamma functions.
The defining RG equations for the evolution of 
these effective quantities are 
\ba
\label{betaf}
Q^2\frac{d}{dQ^2} a(Q^2)&=&\beta(a), \\
\label{gammagf}
Q^2\frac{d}{dQ^2}C_g(Q^2)&=&2 \gamma_g(a)C_g(Q^2), \\
\label{gammaqf}
Q^2\frac{d}{dQ^2}C_q(Q^2)&=&2 \gamma_q(a)C_q(Q^2) \, .
\ea
The RG functions $\beta(a)$ and $\gamma_{g,q}(a)$
describing the evolution of the parameters in the effective scheme 
can be expressed through the $\MSsch$ scheme RG functions
(the standard $\beta$-function of the coupling constant 
and the mass anomalous dimension $\gamma$) using the relations in 
eqs.~(\ref{effdef},\ref{effparam},\ref{effmg},\ref{effmq}).
The explicit formulas of such a RG 
transformation are given in Appendix.
Up to the relative order $\al_s^3$ the RG functions 
$\beta(\al_s)$ and $\gamma(\al_s)$ in the $\MSsch$ scheme 
have been calculated in refs.~\cite{beta4,gamma3c,gamma3v}.
Under the RG transformation 
eqs.~(\ref{effparam},\ref{effmg},\ref{effmq})
the $\MSsch$ scheme $\beta$- and $\gamma$-functions are transformed
into the effective $\beta$- and $\gamma$-functions
(\ref{betaf},\ref{gammagf},\ref{gammaqf}).
Note that first two coefficients
of the $\beta$-function $\beta_0,\beta_1$ and 
first coefficient of the $\gamma$-function
$\gamma_0$ are invariant under RG transformations. 
Numerically the RG functions of the effective parameters describing the 
$\tau$-system up to the order $m_s^2$  are given by
\ba
\label{betafNum}
\beta(a)&=&-a^2\left(2.25 + 4 a + 11.79 a^2 
+ a^3 \left( -76.36 + 4.5 k_3 \right) \right) \, ,\\
\label{gammagfNum}
\gamma_g(a)&=&-a\left(1 + 4.027 a + 17.45 a^2  
+ a^3 \left( 249.59 - k_3  \right)  \right) \, ,\\
\label{gammaqfNum}
\gamma_q(a)&=&-a\left(1 + 4.78 a + 32.99 a^2 
+ a^3 \left(-252.47 - k_3 + 3.38k_{q2}\right) \right) \, .
\ea
For the effective coupling in eq.~(\ref{effdef})
we use the numerical value $a(\mts)=0.1445$ \cite{alpha} 
extracted from the $\tau$ decay rate into nonstrange particles 
within the effective scheme resummation procedure described 
in ref.~\cite{groote}. This value corresponds to 
the $\MSsch$ scheme value of the coupling constant
$\al_s(\mts)=0.343$ that is a bit larger than the most recent result 
obtained in FOPT analysis of the  
$\tau$ decay rate into nonstrange particles \cite{alpha}.  
The PT series for the 
effective $\beta$-function in eq.~(\ref{betafNum}) explicitly 
converges well at the numerical value $a(\mts)=0.1445$.
If the coefficient $k_3$ lies in the range $0<k_3<50$ 
(which is a conservative estimate based on a number of predictions) 
the $a^3$ coefficient in eq.~(\ref{betafNum}) will not be
extremely large but nevertheless the $\beta$-function (\ref{betafNum})
shows asymptotic growth in the N$^3$LO for $k_3>35$.
The $\gamma_g$-function in eq.~(\ref{gammagfNum})
behaves worse than the $\beta$-function but 
still the explicit convergence persists up to the NNLO
at $a(\mts)=0.1445$.
The N$^3$LO correction will show asymptotic growth for values of 
$k_3$ smaller than $129$. 
The $\gamma_q$-function in eq.~(\ref{gammaqfNum}) 
has already shown an asymptotic growth in the 
NNLO which will limit the precision of our results.
The $\beta$- and $\gamma_g$-function will
show asymptotic growth in the N$^3$LO because no
choice of $k_3$ can make them convergent simultaneously. 
This confirms the conclusions of ref.~\cite{onetwo} where the
asymptotic growth for the FOPT 
expressions of the moments of the spectral density
has been found 
in N$^3$LO independently of the choice of the numerical value for $k_3$.
In other words, the evolution of the effective parameters 
for the $\tau$-system is too different
to be handled by the FOPT expressions for the RG functions 
in the N$^3$LO. 

The running of the effective coupling and the mass coefficient 
functions $C_{q,g}(Q^2)$ along the contour in the complex plane 
of momentum squared
is determined by the renormalization group
equations. It is convenient to choose a circular contour
in the complex $Q^2$-plane and to parameterize it
by the relation \mbox{$Q^2=\mts e^{i \phi}$}, \mbox{$-\pi< \phi<\pi$}
which leads to the differential equations
for the coupling constant 
\be 
\label{difa}
- i \frac{d}{d  \phi} a(\phi) = \beta(a(\phi))\, , 
\quad a(\phi = 0) = a(\mts)
\ee
and for the coefficient functions $C_n(\phi)$
\be \label{difm}
- i\frac{d}{d  \phi} C_n(\phi) 
= 2\gamma_n(a(\phi)) C_n(\phi) \, , 
\quad C_n(\phi = 0) = 1
\ee
with $n=q,g$. The solution to the differential equation for 
the running mass in terms of the coefficient functions 
$C_n(\phi)$ in eq.~(\ref{difm}) can be expressed through the integral
\be
\label{expC}
C_n(\phi)=\exp\left(2i\int_0^\phi\gamma_n(a(\chi))d\chi\right).
\ee
The initial values for $a(\phi)$ and $C_n(\phi)$ are fixed 
at $Q^2 = \mts$ or $\phi=0$. 
All corrections stemming from higher order perturbative terms 
are absorbed into the $\beta$- and $\gamma$-function 
coefficients if the effective scheme 
is used as it is defined in eq.~(\ref{effdef}).
The solutions to the differential
equations for the coupling $a(\phi)$ (eq.~(\ref{difa}))
and for the coefficient 
function $C_q(\phi)$ (eq.~(\ref{difm})) 
are shown in Figs.~(\ref{runc},\ref{runmCq}).
The effective coupling $a(\phi)$ does not change 
much uniformly on the contour 
when higher order corrections of the $\beta$-function
are included. Especially the change from NLO to NNLO
is rather small. The behavior of
the coefficient function $C_g(\phi)$ is rather similar to that 
of the effective coupling $a(\phi)$ as one can expect
from the structure of PT series for $\gamma_g(a)$ in
eq.~(\ref{gammagfNum})
in comparison with the $\beta$-function in eq.~(\ref{betafNum}).
The function $C_g(\phi)$ appears to converge uniformly in the
interval $|\phi|<\pi$ when going from NLO to NNLO. 
The function $C_q(\phi)$ (Fig.~(\ref{runmCq})) 
does not converge uniformly on the 
contour because of the slow convergence of the 
$\gamma_q$-function (eq.~(\ref{gammaqfNum})) in consecutive orders
of PT expansion.   
  \begin{center}
\begin{figure}[!ht]
\label{fig1b}
       \epsfig{file=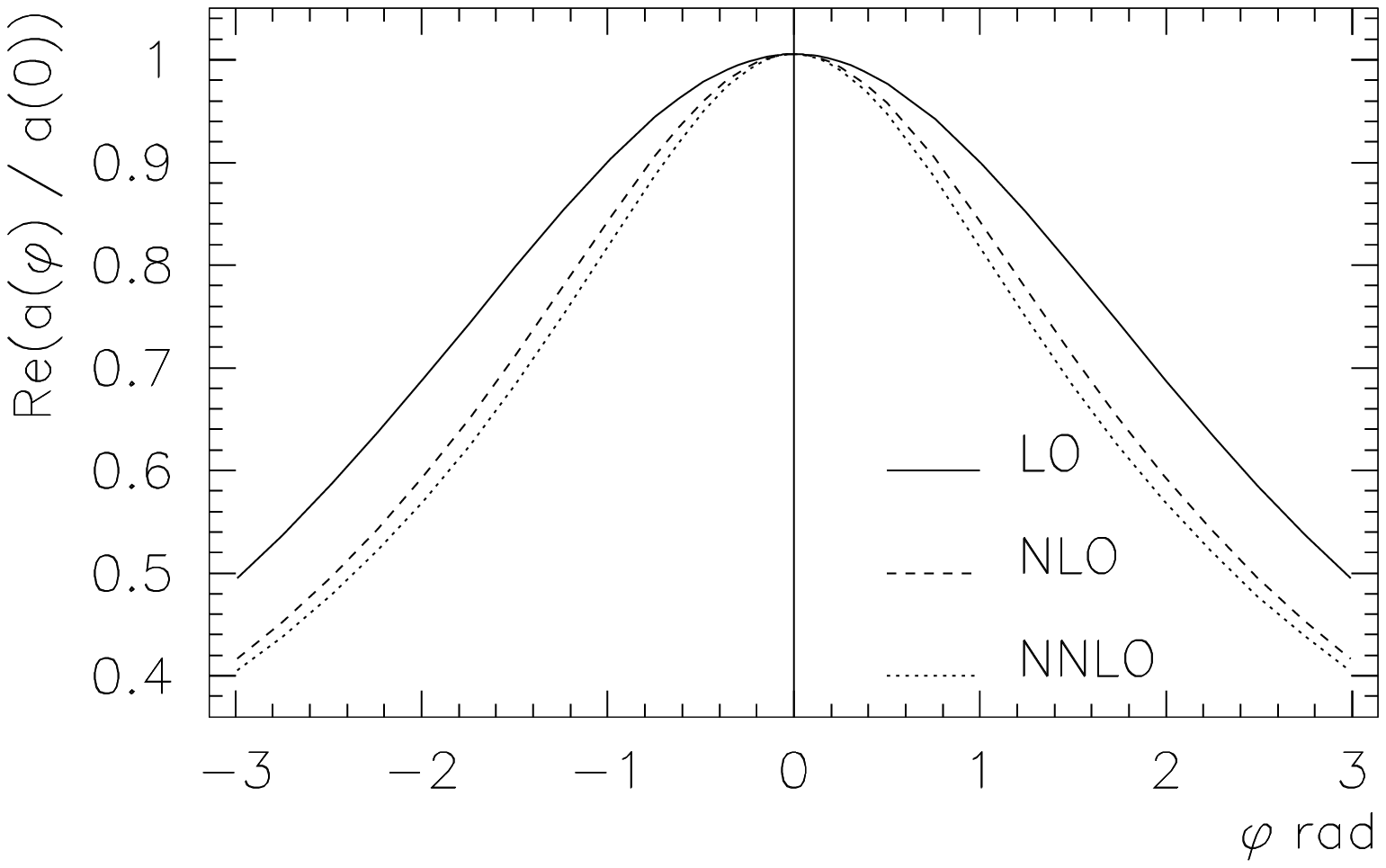,angle=0,width=.45
                 \textwidth,height=.33\textwidth}
       \quad \quad
       \epsfig{file=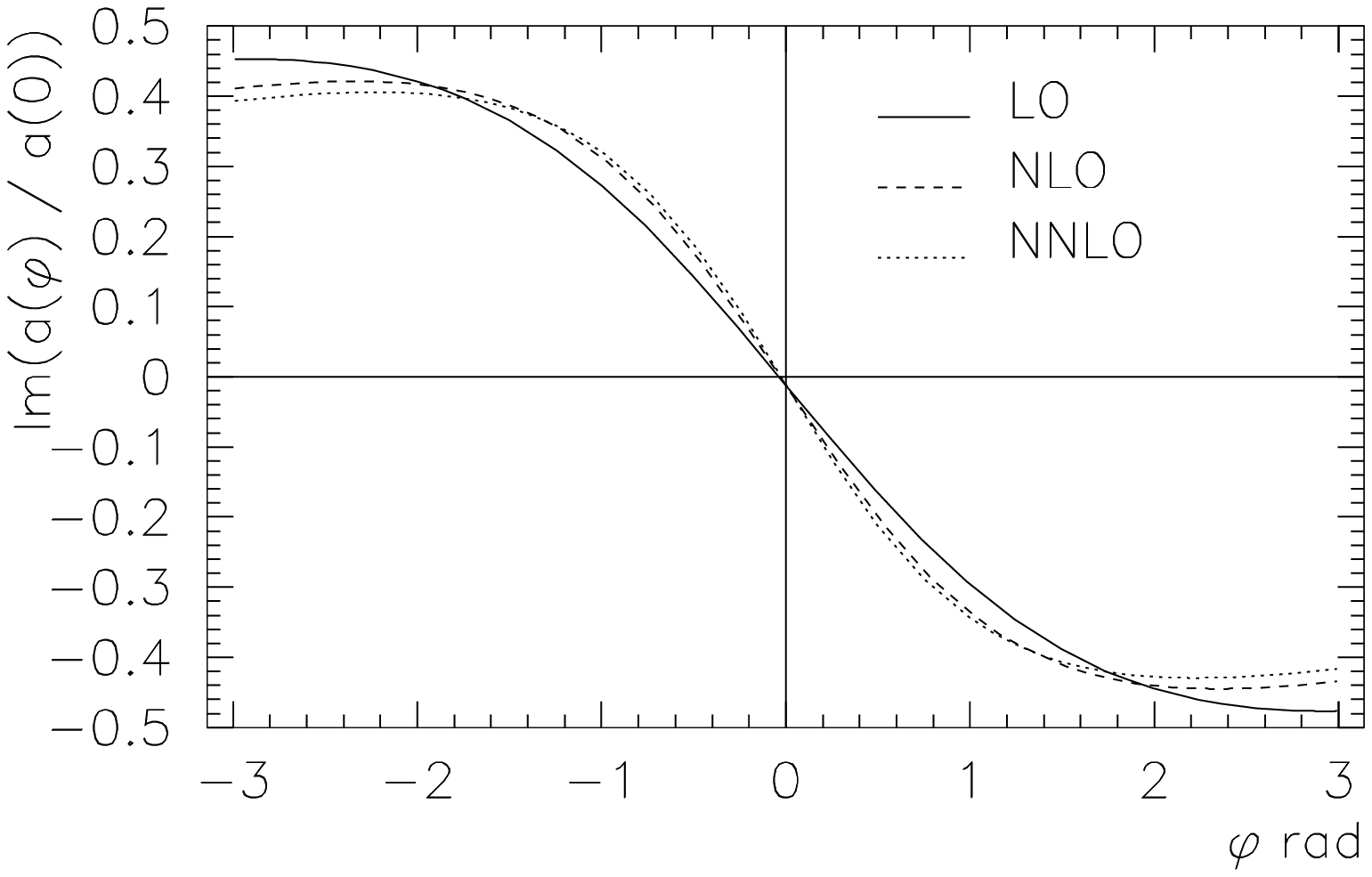,angle=0,width=.45
                  \textwidth,height=.33\textwidth}
       \caption{\label{runc}
Running of the effective coupling $a(\phi)$ on a circular contour 
in the complex plane calculated at LO, NLO and NNLO 
(left: real part; right: imaginary part)}
    \end{figure}
\begin{figure}[!ht]
       \epsfig{file=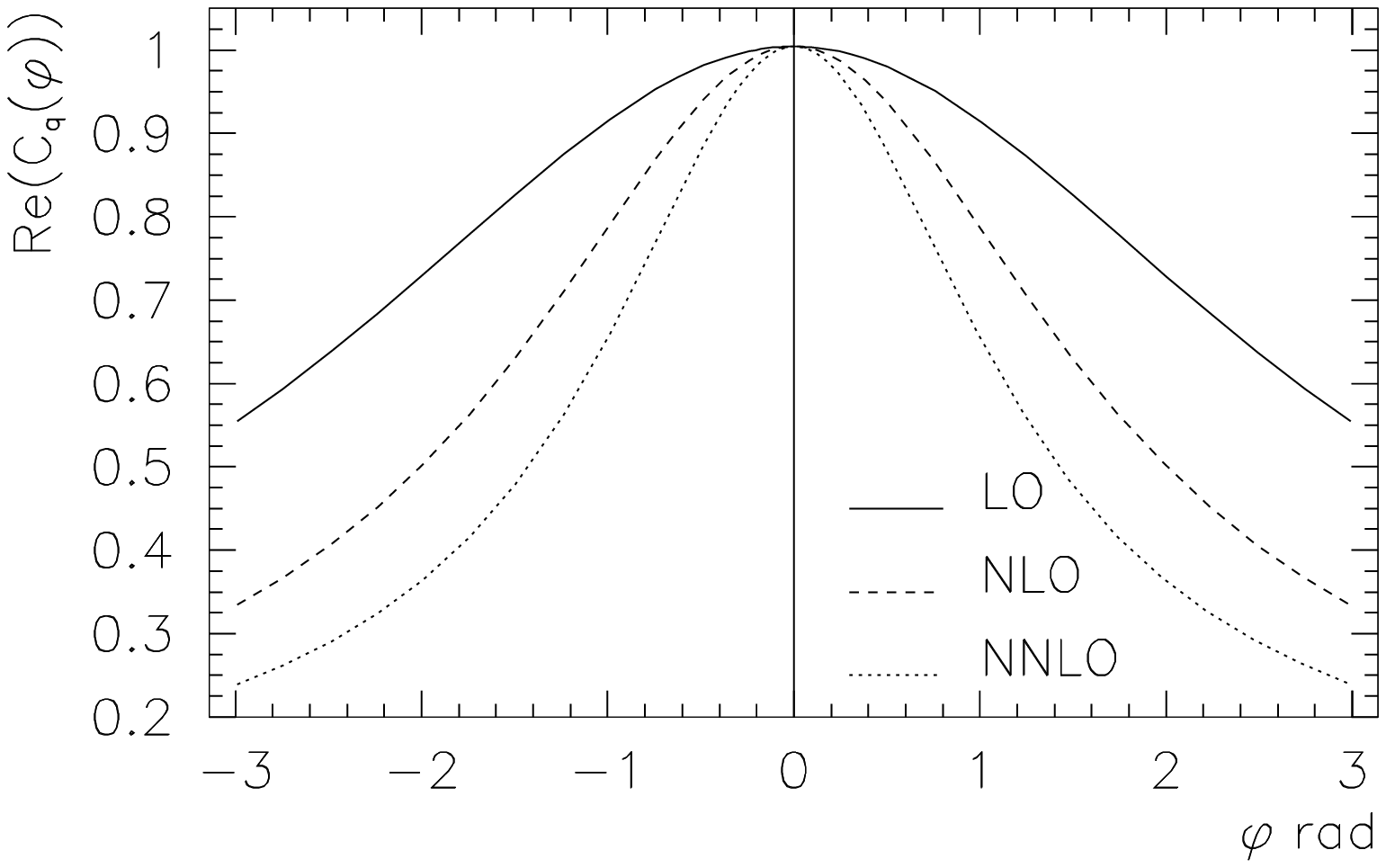,angle=0,width=.45
       \textwidth,height=.33\textwidth}
       \quad \quad
       \epsfig{file=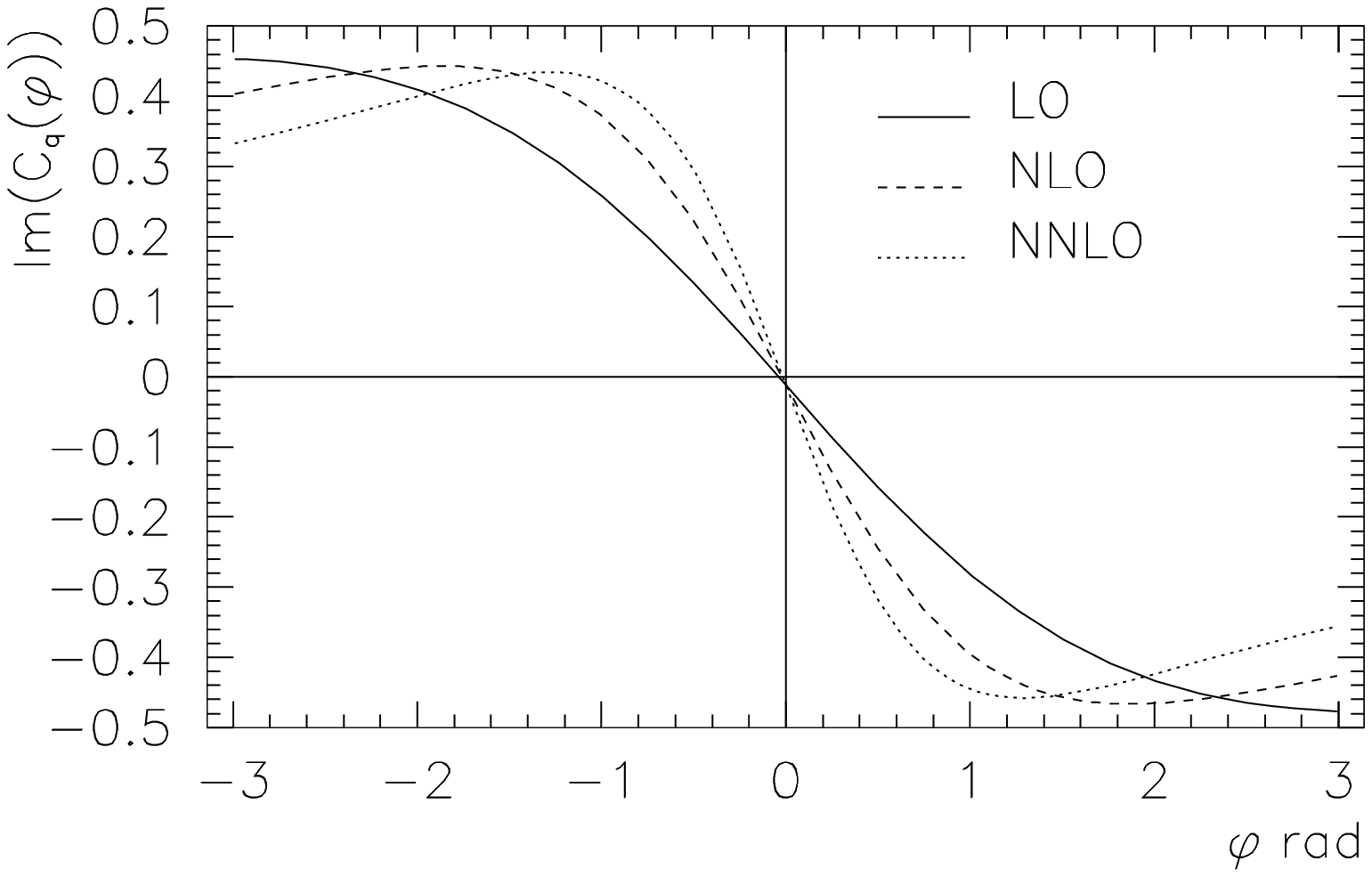,angle=0,width=.45
       \textwidth,height=.33\textwidth}
       \caption{\label{runmCq}
Running of the mass coefficient function $C_q(\phi)$ on a circular 
contour in the complex plane calculated at LO, NLO and NNLO
(left: real part; right: imaginary part)}
\end{figure}
\end{center}     

\subsection{Resummation} 
\label{Resummation}
We use the direct generalization of the resummation procedure described 
for the massless case in ref.~\cite{groote}. 
We treat the renormalization group functions $\beta(a)$ and $\gamma_n(a)$
in any fixed order of PT (LO, NLO and NNLO)
as exact functions and solve exactly (though numerically) the 
differential equations~(\ref{difa},\ref{difm}) for the running 
parameters $a(\phi)$ and $C_{q,g}(\phi)$ on a circular 
contour in the complex plane. 

Having the solutions for the running parameters we calculate 
numerical values for the theoretical expressions of the observables in
the $\tau$-system. The observables of interest are expressed through
the moments of the spectral density of the basic correlator
given in eq.~(\ref{corrF}).
For the massless part of the correlator $\Pi(q^2)$
given in eqs.~(\ref{expmq},\ref{expmg}) we define 
moments of the spectral density by the relation
\ba
\label{masslessmom}
M(n) &=& (n+1) \frac{i}{2 \pi} 
\oint \Pi(q^2) 
\left(\frac{q^2}{\mts}\right)^n 
\frac{dq^2}{\mts} \nn \\
&=&1+(-1)^n\frac{1}{2\pi}
\int_{-\pi}^{\pi}e^{i(n+1)\phi} a(\phi) d\phi 
+ \frac{1}{2\pi}\int_{-\pi}^{\pi} a(\phi) d\phi \, . 
\ea
For the mass correction related to the $g$-part of the correlator
the expressions for the moments are very similar. One has 
\be\label{gmoments}
M_g(n)=\frac{(-1)^n}{2\pi}\int_{-\pi}^{\pi} e^{i (n+1)\phi} 
C_g(\phi)d\phi + \frac{1}{2\pi}\int_{-\pi}^{+\pi}C_g(\phi)d\phi \, .
\ee 
For the mass correction in the $q$-part of the correlator
we use the definition of physical moments given in \cite{onetwo}. 
The corresponding expression on the contour reads
\be
\label{qmompart}
M_q^{ph}(n)=(-1)^{n+1}\frac{1}{2\pi}\int_{-\pi}^{\pi} 
C_q(\phi) e^{i n \phi} d\phi \, .
\ee
The moments of the massless part $M(n)$  
and the mass corrections $M_{q,g}(n)$  are the basic objects 
which can be calculated theoretically. 
Our qualitative conclusions about the stability  
of the running parameters due to the 
higher order corrections to the $\beta$- and $\gamma$-functions are
confirmed by the behavior of the moments. 
The moments of the massless spectral density 
given in eq.~(\ref{masslessmom}) are stable with regard to inclusion
of higher order corrections of the $\beta$-function. 
This is a direct consequence of the $\beta$-function behavior
in consecutive orders of PT 
along the integration contour plotted in Fig.~\ref{runc}. 
The moments of the $g$-part given in 
eq.~(\ref{gmoments}) are less stable with regard to inclusion
of higher order corrections of the $\gamma_g$-function
than the moments of the massless part.
The moments of the $q$-part in eq.~(\ref{qmompart})
behave worse in higher orders than the moments of the $g$-part 
which reflects the slower convergence of the $\gamma_q$-function
in eq.~(\ref{gammaqfNum}) as compared to $\gamma_g$-function
in eq.~(\ref{gammagfNum}).
In the massless and in the $g$-part the values of the 
moments are dominated by the second integral in 
eqs.~(\ref{masslessmom},\ref{gmoments})
because the values of the integrals with oscillating integrands 
in eqs.~(\ref{masslessmom},\ref{gmoments}) 
decrease for high order moments. 
The coefficient functions $C_{q,g}(Q^2)$
depend on the $\gamma_{q,g}$-functions exponentially 
according to eq.~(\ref{expC}) and they are 
very sensitive to the convergence of 
the $\gamma_{q,g}$-functions.
The exponential dependence of the mass coefficient functions 
on RG functions is 
an essentially new feature of the analysis of the $m_s^2$ corrections
as compared to that of the massless part of the $\tau$ lepton decay rate.
Such strong dependence requires more accurate treatment 
of PT series for RG functions.

\section{Determination of $m_s$ from $\tau$ decays}
One of the important aims of the analysis of 
Cabibbo-suppressed $\tau$ decays is the 
extraction of a numerical value for the mass parameter $m_s$. 
Various observables can be used for this purpose.
Here we consider the $\tau$ lepton total decay rate
into hadrons.
Experimental data for Cabibbo-suppressed 
hadronic $\tau$ decays are not yet very precise.
Contrary to the experimental data the theoretical expressions 
for the relevant observables related to the moments 
of the spectral density of the correlator of the hadronic currents 
are known rather precisely in a sense that 
the corresponding PT series within OPE are 
calculated in high (next-to-next-to-leading) orders
of expansion in the strong coupling constant.
However, this theoretical accuracy is quite formal, because 
the explicit convergence of PT expressions for the moments is rather 
slow and the PT series has to be resummed in order to obtain
numerical results.

The theoretical expression for the $m_s^2$ corrections to 
moments $(k,l)$ of the differential decay rate 
is given by the contour integral in the complex $q^2$-plane
\ba
\label{fincor}
R_{m\tau}^{kl} &=& \frac{i}{2 \pi} 
\oint 2 \left( 1- \frac{q^2}{\mts} \right)^{2+k} 
\left( \frac{q^2}{\mts} \right)^l 
3\left(\frac{m_s^2 \Pi_{mq}(q^2)}{q^2} 
- \frac{m_s^2}{\mts}
\Pi_{mg}(q^2)  \right)\frac{dq^2}{\mts} \nn \\
&=& -6\left(\frac{m_q^2}{M_\tau^2} A_{kl} 
     + \frac{m_g^2}{M_\tau^2}
B_{kl}\right) 
=
-6\frac{m_s^2}{M_\tau^2} 
\left(\omega_q A_{kl} + \omega_g B_{kl} \right) 
= -6\frac{m_s^2}{\mts} F_{kl} 
\ea
where the superscript $(k,l)$ denotes 
an integration with additional 
weight factors which suppress the high 
$(k>0,l=0)$ and low $(k=0,l>0)$ energy region.  
The finite order PT expressions for the coefficients
$F_{kl}$ can be found in refs.~\cite{onetwo,msNPB}.
The analyses performed in ref.~\cite{onetwo,msNPB}
in FOPT and CIPT within resummation in the $\MSsch$ scheme
have demonstrated that 
the explicit convergence of the PT series for the moments is slow.
A close conclusion about the convergence of PT series in
the $\MSsch$ scheme was given in ref.~\cite{pichprades}.
Therefore, in the present paper we use the contour resummation
procedure in the effective scheme 
and calculate the relevant coefficients $F_{kl}$
in a closed form. 

We use the $(0,0)$ moment as our best 
theoretical estimate for comparison with experiment.
Other moments are briefly discussed but 
are not used in numerical analysis for 
the strange quark mass determination.
The choice of optimal moments is discussed in detail later.
In brief, the reasoning is based on analyzing which energy
regions saturate the respective moments. Indeed, 
to choose a special linear combination 
of moments is equivalent to integrating the running parameters 
with a special weight function. Whether the corrections 
to a given linear combination are large or small can depend strongly
on the integration region in which the integral 
of the spectral density is saturated. 
The change of the running parameters with
higher order corrections is big in the region close to 
the physical cut (at $\phi= \pm\pi$) and 
small in the deep Euclidean region where the functions 
are fixed by their starting values.
If some observables have the regions 
$\phi=\pm\pi$ strongly suppressed they have no contribution from 
the perturbative running and, therefore, are very stable with
regard to the inclusion of higher order terms of RG functions.
The moments of the type $(k,0)$ represent such observables
for large $k$. However, these moments have the contributions from
large energy region strongly suppressed and 
receive relatively large contributions from the low energy
region that makes them rather nonperturbative. 
Therefore, the use of the moments $(k,0)$ with large 
values of $k$ is not under reliable quantitative control within OPE 
and perturbation theory, though their use is favorable from the point 
of view of precision of experimental data.
We elaborate on this point later.

\subsection{Relation of 
the $\MSsch$ mass to effective mass parameters}
Within the effective scheme 
approach the $\tau$-system is described in its own terms
with the mass parameters $m_{q,g}$.
In order to obtain a 
numerical value for $m_s$ which can be compared with other
determinations
we express the natural mass parameters of the $\tau$-system $m_{q,g}$ 
through the $\MSsch$ mass parameter $m_s$.
We emphasize that this is only done for purposes of comparison.
In principle, observables in the 
$\tau$-system are best described by their  
internal parameters $a, m_q$ and $m_g$. 
Relations between
observables within the $\tau$-system can be found 
without any reference to the standard $\MSsch$
scheme parameters.

In general, the perturbation theory expression for a given
observable in a given 
renormalization scheme is parametrized by a mass scale and by
coefficients of the evolution (RG) functions.
In the massless case these are the scale parameter $\Lambda$ 
and the perturbation theory coefficients of the 
$\beta$-function (e.g. \cite{stevenson,stevenson1}).
In the massive case there are in addition the invariant mass $M$ 
to be defined in eq.~(\ref{invmass}) and 
the coefficients of the $\gamma$-function
describing the evolution of the running mass. 
As in the case of the scale parameter $\Lambda$, 
the invariant mass $M$ can be defined in different ways.
The concrete definition may be fixed by a given 
asymptotic behavior at large momenta. This is the way 
the standard scale parameter
$\Lambda_{\MSsch}$ is fixed. We define the invariant mass 
$M$ in a $\mu$ independent way by writing
\be 
\label{invmass}
M = \frac{m(\mu^2)}{a(\mu^2)^{\gamma_0/\beta_0}} 
\exp \Bigg\{- \int_0^{a(\mu^2)} 
\left( \frac{\gamma(\xi)}{\beta(\xi)} -
       \frac{\gamma_0}{\beta_0 \xi} 
\right) d \xi \Bigg\} \;.  
\ee
Note that $M$ is renormalization group invariant. 
That means that if $m(\mu^2)$ is redefined 
by some RG transformation the change
is absorbed by the corresponding change of the  
$\gamma$-function so that $M$
remains invariant up to the order in the coupling 
which has been taken into account.
The RG invariance of $M$ can be used to relate 
running masses defined in different renormalization 
schemes. After squaring eq.~(\ref{invmass}) 
we find the relation between two mass definitions in 
different schemes with the $\gamma$-functions
$\gamma(a)$ and $\gamma^\prime(a)$ expressed in
terms of the same coupling $a$. In our analysis we use the 
effective coupling $a$ as it is defined in eq.~(\ref{effdef})
and relate $m_n^2$ ($n=q,g$) to ${m'}_s^2$ by
\be 
\label{relmass}
m_n^{2}(\mu^2) = {m'}_s^2(\mu^2) 
\exp \Bigg\{ -2 \int_0^{a(\mu^2)} 
\frac{\gamma'(\xi)-\gamma_n(\xi)}{\beta(\xi)} d\xi  
\Bigg\} \;. 
\ee
Here ${m'}_s^2$ is the standard $\MSsch$ scheme mass but with 
the evolution function $\gamma'(a)$ expressed 
through the effective coupling. Eq.~(\ref{relmass})
relates two mass parameters defined in different renormalization
schemes but expressed through 
the same coupling. Therefore the standard $\gamma$-function
of the $\MSsch$ mass should be rewritten 
in terms of the effective parameter 
$a$ before the use in eq.~(\ref{relmass}). 
It is also possible to use eq.~(\ref{invmass}) 
written in the $\MSsch$ scheme and in the effective scheme.
Then one determines the 
relation between $m_{q,g}$ and $m_s$ by eliminating 
the invariant mass $M$ directly
\ba
\label{relmass2}
\frac{m_n^2(\mu^2)}{m_s^2(\mu^2)}&=&
\left(\frac{a_{n}(\mu^2)}{a_{\MSsch}(\mu^2)}
\right)^{\frac{2 \gamma_0}{\beta_0}}
\times \exp \Bigg\{-2\int_0^{a_{\MSsch}(\mu^2)} 
\left(\frac{\gamma_{\MSsch}(\xi)}{\beta_{\MSsch}(\xi)} 
- \frac{\gamma_0}{\beta_0\xi} \right) d\xi \Bigg\} \nn \\ 
&& \times \exp \Bigg\{ 2 \int_0^{a_{ n}(\mu^2)} 
\left(\frac{\gamma_{n}(\xi)}{\beta_{\rm{eff}}(\xi)} 
- \frac{\gamma_0}{\beta_0\xi}\right) d\xi \Bigg\}\, .
\ea         
In our case the effective couplings $a_n$, $n=q,g$ 
are equal to the same coupling $a$ for both masses $m_{q,g}$.
The two procedures described above 
lead to close numerical values for the coefficients
$\omega_{q,g}$
relating the effective mass parameters $m_{q,g}$ to 
the $\MSsch$ scheme mass $m_s$.
The difference of numerical values for $\omega_{q}$ and $\omega_g$
obtained from the two procedures 
described by eqs.~(\ref{relmass}) and (\ref{relmass2}) 
turns out to be less
than $3\%$ which is the residual scheme dependence of the results. 
Eqs.~(\ref{relmass},\ref{relmass2}) allow us to express
the internal parameters $m_{q,g}$ through the standard
$\MSsch$ scheme parameter $m_s$
(see eqs.~(\ref{effmg},\ref{effmq}) and (\ref{gammagf},\ref{gammaqf})). 
Finally we find for the coefficients $\omega_{q,g}$
relating the effective mass parameters $m_{q,g}$
to the reference $\MSsch$ mass $m_s$
\ba
\label{ome}
m_q &=&\omega_q m_s\, , \quad m_g=\omega_g m_s \,, \nn \\
\omega_q &=&  1.73 \pm 0.04 , \quad \omega_g = 1.42 \pm 0.03\;.
\ea
The numerical values for the coefficients $\omega_{q,g}$ are not close
to unity which shows that perturbation theory corrections 
for observables in the $\tau$-system are 
rather large in the $\MSsch$ scheme.
The FOPT expansion for the coefficients $\omega_{q,g}$  
has been obtained in ref.~\cite{onetwo}. 
This expansion converges slowly
which forces us to use the exact RG conversion 
given in eqs.~(\ref{relmass},\ref{relmass2}).

\subsection{Power corrections from dimension $D=4$ \\
condensate terms}
For the determination of $m_s$ one needs not all $D=4$ condensate
corrections to the theoretical expression for the 
$\tau$ lepton decay rate but only those that enter 
the difference
\be
\delta \! R_\tau^{kl} 
=\frac{R_{\tau s=0}^{kl}}{|V_{ud}|^2} 
-\frac{R^{kl}_{\tau s=1}}{|V_{us}|^2}\, .
\ee
Here $R_{\tau s=0,1}^{kl}$ is defined as
\be
R_{\tau s=0,1}^{kl} 
= \int_0^{\mts} ds \left(1- \frac{s}{\mts} \right)^k
\left( \frac{s}{\mts} \right)^l 
\frac{d R_{\tau s = 0,1}}{ds} 
\ee
and $dR_{\tau s = 0,1}/ds$ is the 
differential $\tau$-decay rate into
the final hadronic states with the strangeness $0,1$ and 
the energy $\sqrt{s}$. 
In the theoretical expression for the 
difference $\delta \! R_\tau^{kl}$ we neglect terms of 
the order $m_s^3/M_\tau^3$, set the $u$- and $d$-quark masses to zero, 
and retain only the most important term linear in $m_s$
(cf. refs.~\cite{msNPB,pichprades}).
Within OPE the coefficient of this term is given by the quark
condensate. The final result reads
\be \label{finelres}
\delta \! R_\tau^{kl} = N_c S_{EW} 
\left(6\frac{m_s^2}{M_\tau^2} F_{kl} 
- 4\pi^2\frac{m_s}{M_\tau}\frac{\langle \bar s s \rangle}{M_\tau^3} 
T_{kl} \right)
\ee
where $N_c=3$, $m_s\equiv m_s(\mts)$, and $S_{EW} = 1.0194$ 
describes electroweak corrections \cite{ewcorr1,ewcorr2}.
The coefficient function of the $D=4$ local operator 
$m_s\bar s s$ is taken in the leading 
order of perturbation theory expansion within OPE.
In this approximation for the coefficient function 
the quantities $T_{kl}$ multiplying the quark condensate
are given by the expression
\be
T_{kl} = 2\left( \delta_{l,0} (k+2) - \delta_{l,1} \right)\, .
\ee
The numerical values for the first few coefficients $T_{kl}$ 
read
\be
T_{00} = 4, \quad T_{10}  = 6, \quad
T_{20} = 8,  \quad  T_{01} = -2, \quad T_{11} = -2\, .  
\ee
These results agree with the leading order expressions for the
coefficients given in ref.~\cite{pichprades}.
We have set up all ingredients necessary for the 
evaluation of eq.~(\ref{finelres}). Table~\ref{T1} gives 
the coefficients
of the $q$- and $g$-mass parameters in eq.~(\ref{fincor}). 

\section{Numerical analysis and the choice of moments}
Having theoretical expressions for all moments we have still 
to optimize the choice for comparison with experiment and
extraction of the numerical value for the strange quark mass.
As was mentioned above the theoretical expressions for 
the moments $(0,l)$, $l>0$ are more reliable
from the point of view of perturbation theory than those for 
the moments $(k,l)$ with nonzero $k$ (the detailed discussion
of this point is given below).
However, experimental precision is worse for the moments 
$(0,l)$ with large $l$ because such moments are saturated by
the contributions of many-particles hadronic states which is difficult
to measure (see Table~\ref{T2}) . Note that some 
many-particles hadronic state contributions in the experimental
data (for instance, $K4\pi$ contribution) 
are represented by a result of Monte-Carlo simulation
rather than direct measurements. Therefore, we use the moment $(0,0)$
as our best choice from both experimental and theoretical point of
view. The theoretical expression for this moment 
exhibits a rather good convergence
in consecutive orders of perturbation theory
and the accuracy of experimental data for it is still acceptable
in comparison with higher $(0,l)$ moments.
Note that the perturbative convergence is the main concern 
of the theoretical analysis in both massless and massive cases.
In the massless case the nonperturbative corrections are small 
if factorization is used for the four-quark condensates
\cite{BraNarPic92,SVZ,factor}. 

\subsection{Numerical value for the strange quark mass}
The coefficients $A_{kl}$ are related to the $q$-part of the correlator.
This part contains contributions from spin 0 and spin 1 particles.
The spin 0 piece is prone to possible 
nonperturbative contributions of (direct)
instantons and perturbation theory expansions are expected to 
break down in low orders in this channel.
However, this is only a general expectation without strict 
quantitative estimates of applicability of PT.
The coefficients $B_{kl}$ are related to the $g$-part of the
correlator. The $g$-part contains only contributions of 
spin 1 particles. Nonperturbative contributions of (direct)
instantons are forbidden in this channel by symmetry 
considerations. 
\begin{table}
\[ 
\begin{array}{||c||c|c|c||c|c|c||}
\hline \hline
(k,l) & A_{kl}^{\rm{LO}} & A_{kl}^{\rm{NLO}}  
& A_{kl}^{\rm{NNLO}} & B_{kl}^{\rm{LO}} & B_{kl}^{\rm{NLO}}  
& B_{kl}^{\rm{NNLO}}   \nn \\ \hline
(0,0) & 1.361  & 1.445  & 1.434  &0.523 & 0.601  & 0.625 \\
(1,0) & 1.568  & 1.843  & 1.976  &0.441 & 0.552  & 0.601 \\
(2,0) & 1.762  & 2.282  & 2.646  &0.390 & 0.530  & 0.607 \\ \hline  
(0,1) &-0.207  & -0.398 &-0.542  &0.082 & 0.050  & 0.025 \\  
\hline \hline 
\end{array}
\]
\caption{\label{T1}Coefficients of eq.~(\ref{fincor})}
\end{table}
The coefficient $F_{kl}$ of $m_s^2$ term in 
eq.~(\ref{finelres}) is combined according to eq.~(\ref{fincor})
\be
\label{abfdef}
F_{kl} =\omega_q A_{kl} + \omega_g B_{kl} .
\ee
As for the linear in $m_s$ term in eq.~(\ref{finelres}) 
we calculate its numerical coefficient using 
a phenomenological value for the quark condensate. 
We use the relation \cite{narga,gammass,newgam}
\be
\label{stqucond}
\langle \bar s s \rangle 
= (0. 8 \pm 0.2 ) \langle \bar u u \rangle 
\ee 
and the numerical value
$\langle \bar u u \rangle = - (0.23~\rm{GeV} )^3$
which coincides with the standard value (see e.g.~\cite{pichprades}).
Substituting all necessary quantities into eq.~(\ref{finelres})
we arrive at the defining equation for $X=m_s/(130~{\rm MeV})$
\be
\label{finaleqms}
\frac{1}{N_c S_{EW}}\left(\frac{M_\tau}{130~{\rm MeV}}\right)^2
\delta \! R_\tau^{kl} 
= X \left(6 F_{kl} \cdot X + 0.936 \cdot T_{kl}\right)\, .
\ee
The dimension-four term contributes appreciably 
to the total theoretical result
for different moments. For the moments $(k,0)$ its relative
contribution increases with $k$ for the first few moments.
We use only the moment $(0,0)$ for which the dimension-four
contribution gives about 16\% of the total result.
The coefficient function of the dimension-four contribution converges
well in the perturbative expansion within OPE. 
We do not take the PT corrections to the coefficient function of the 
dimension-four contribution into account (see e.g.~\cite{pichprades})
because of the large
uncertainty in the numerical value of the strange quark condensate
in eq.~(\ref{stqucond}).

For extraction of the numerical value for the strange quark mass
we use the experimental data obtained by the ALEPH collaboration 
\cite{exp1}. The results for $m_s(\mts)$
from different moments $\delta \! R_\tau^{kl}$
are given in Table~\ref{T2}.
\begin{table}
\[
\begin{array}{|c|c|c|}
\hline
(k,l)&(\delta\! R_\tau^{kl})^{\rm exp}& m_s(\mts)~{\rm MeV}\\ \hline
(0,0)& 0.394 \pm 0.137 & 130 \\
(1,0)& 0.383 \pm 0.078 & 111 \\
(2,0)& 0.373 \pm 0.054 & 95 \\ \hline
\end{array}
\]
\caption{\label{T2} Results for $m_s(M_\tau)$ 
obtained from different moments of $\delta \! R_\tau$}
\end{table}
For the determination of $m_s$  
we use only the moment $(0,0)$ as the most reliable one from
the perturbation theory point of view.
A detailed discussion of 
the justification for our choice is given later.

The final relation for determining 
the central value of the strange quark mass 
from the data on Cabibbo-suppressed $\tau$ decays reads 
\be
\label{finaleqms00}
\frac{1}{N_c S_{EW} }\left(\frac{M_\tau}{130~{\rm MeV}}\right)^2
(\delta \! R_\tau^{00})^{\rm exp} 
= 24.1 = X \left(20.2 X + 3.74 \right)\, .
\ee
The result is $X=m_s/(130~{\rm MeV})=1.00\ldots$ 
with the accuracy of two decimal places.
This leads to our final prediction for the strange 
quark mass at $M_\tau$
\be 
\label{fipred}
m_s(\mts)=130 \pm 27_{\rm{exp}} \pm 3_{\langle\bar{s}s \rangle} 
\pm 6_{\rm{th}}~{\rm MeV}\, .
\ee
Note that this result is obtained with 
the numerical value for the effective coupling $a(\mts)=0.1445$
extracted from the $\tau$ decay rate into nonstrange particles 
within the effective scheme resummation procedure described 
in ref.~\cite{groote}. The reference numerical value 
of the $\MSsch$ scheme coupling constant is $\al_s(\mts)=0.343$.

We also give the value for $m_s(1~{\rm GeV})$ obtained from
$m_s(\mts)$ in eq.~(\ref{fipred}) after 
four-loop running in the $\MSsch$ scheme 
\be 
\label{mt1eff}
m_s(1~{\rm GeV})=176 \pm 37_{\rm exp} 
\pm 4_{\langle\bar{s}s \rangle} 
\pm 9_{\rm th}~{\rm MeV} \, .   
\ee
The numerical value of the strange quark mass at the scale 
$1~{\rm GeV}$ depends on the way the evolution is performed
because of the truncation of the PT series for the RG functions.
This difference reflects the residual scheme dependence of the
evolution.

The numerical value for the invariant mass $M$ 
defined in eq.~(\ref{invmass}) and calculated in the $\MSsch$ scheme 
with $\al_s(\mts)=0.343$ reads
\be
\label{invMnum}
M=312 \pm 65_{\rm{exp}} \pm 7_{\langle\bar{s}s \rangle} 
\pm 14_{\rm{th}}~\rm{MeV} \, .
\ee
This value can be used for comparison of the results 
of the strange quark mass determination obtained from different 
theoretical calculations and experimental data. 
If the effective scheme is used for determination of
the numerical value for the invariant mass $M$ from
eq.~(\ref{invmass}) the result is slightly different that 
reflects the residual scheme dependence due to truncation of PT series
for $\beta$- and $\gamma$-functions. 

Results presented in eqs.~(\ref{mt1eff},\ref{fipred})
are the main new numerical results of
the paper. Another new results is a formulation of the problem
in an effective scheme and the development of all the necessary
techniques for phenomenological applications.
 
Note that the uncertainty of the final result is smaller than
the difference between the results of mass extraction from the
zero and second moment 
(first and the last lines of Table~\ref{T2}).
As we have mentioned before and explain in detail later on
we do not consider the results obtained from the high moments
as reliable which means that the theoretical uncertainties of the numerical
mass value $m_s^{(1)}=111~\rm MeV$ (from the first moment) and
$m_s^{(2)}=95~\rm MeV$ (from the second moment)
are expected to be much larger than the uncertainty of our
optimal choice -- the zeroth moment.
Therefore, the results of all determinations 
are consistent but the uncertainty of higher moments is
large. Our view is that 
after accounting for high dimension condensates the second
and third determination will change (while the first one does
not) and may lie within the error bars.
Since there is no reason to expect
the effect of high-dimension condensates to be negligible
the requirement that the first and third determination
are the same implicitly implies that 
high-dimension condensates are negligible which is just an
additional assumption (without any justification).
Because we do not use these moments in our determination we do
not quantitatively discuss these uncertainties.

\subsection{Comparison with other results}
Our final result for the numerical coefficient 
in front of the $m_s^2$ correction in eq.~(\ref{finaleqms00})
agrees with the estimate obtained from the analysis in
the $\MSsch$ scheme \cite{msNPB}.
In the present paper we find $20.2\pm 1.8$ for the 
coefficient of the $m_s^2$ correction
while the final result of ref.~\cite{msNPB}
with the conservative estimate of the error bar is $18.1\pm 4.8$.
The final uncertainty of the result obtained in 
ref.~\cite{msNPB} was determined from both
resummed and FOPT analyses. If only the CIPT analysis in 
the $\MSsch$ scheme 
is used then one obtains the value $18.1\pm 2.6$ which 
has a smaller uncertainty \cite{msNPB}.
The present value $20.2\pm 1.8$ results from the analysis
in the effective scheme with stricter criteria of convergence
and more conservative error estimate;
for instance, in the final result for $F_{kl}$
we have doubled the error for coefficients $\omega_{q,g}$
from eqs.~(\ref{ome}).
However, the interpretation of the perturbation series in 
a closed manner (resummation in the effective scheme) 
allows one to reduce the uncertainty 
of theoretical expressions in comparison with 
the previous analysis in the $\MSsch$ scheme performed 
in ref.~\cite{msNPB}.
This leads to an essential reduction of the theoretical 
part of the error in the extracted numerical value for the
strange quark mass. 

The analysis of mass corrections in the $\MSsch$ scheme was also done
in ref.~\cite{pichprades} where a very accurate account for known
PT corrections to the coefficient functions of nonPT
corrections was given.
The direct comparison of our results with
ref.~\cite{pichprades} is not simple because different 
representations have been chosen for the observables.  
Also some approximations have been used which prevents
us from directly comparing with the results of the present paper 
at intermediate stages. The
final results for the mass extracted from the zeroth order
moment is rather close
to the result of ref.~\cite{pichprades}. Still in general the
results of refs.~\cite{msNPB,pichprades} are rather close
to each other concerning use of scheme while the change of the scheme was the
main reason for our analysis as for comparison 
with ref.~\cite{msNPB}. The analysis of the present paper 
also extends the analysis of
scale dependence which has been done in ref.~\cite{pichprades}.

The resulting values of the strange quark mass
are close to the earlier estimates obtained in 
refs.~\cite{narma,gassLeut,oldmass,oldmass1}
with the use of less accurate theoretical input formulas
and less precise experimental data.
The recent analysis based on (pseudo)scalar correlators 
with a thorough parameterization of experimental data
gives a value close to ours \cite{recent1,recent2}.
In a lattice calculation of the numerical value for the 
strange quark mass \cite{lattice1,lattice2}
the theoretical input is of nonperturbative nature. The recent results
obtained in lattice calculations are smaller than our value for $m_s$
but still inside the error bars as given in eq.~(\ref{fipred}).

\subsection{Choice of moments}
High moments with the weight function $(1-s/M_\tau^2)^k$
(large values of $k$)
are saturated by nonperturbative (infrared, or low-energy) contributions
because the perturbative region of integration (large energy)
is suppressed. 
On the experimental side this means
that such moments are saturated by the contributions of low lying
resonances. In the considered case this is the contribution 
of the $K$ meson in spin 0 channel and of the system 
$K_1(1270)-K_1(1400)$ in the spin 1 channel. The low-energy contributions
can be accurately measured that makes the
moments obtained from the experimental data rather precise.
On the theoretical side, within the OPE calculation of the correlator
the infrared sensitivity of high-$k$ moments means 
that the PT contribution to such moments is suppressed 
and the moments are saturated by the contributions of vacuum condensates
of high-dimension operators. 
Implicitly this is seen in a poor convergence of the PT series for
such moments which is demonstrated in \cite{onetwo}. 
To obtain an accurate numerical value for such
moments one has to include vacuum condensates of 
high-dimension operators. However, numerical values for 
vacuum condensates of the local operators with dimension 
larger than six are completely unknown
and are usually neglected in the $\tau$-system analyses.
Therefore, high moments with the weight function $(1-s/M_\tau^2)^k$
have an uncontrollable admixture of high-dimension
condensates that makes them strongly nonperturbative
and, therefore, unreliable for practical PT applications.
Indeed, the moments $(k,0)$ obtain contributions from all condensates 
up to dimension $D=2k+8$ the numerical values of 
which are unknown for large $k$.
Because on the experimental side this contribution corresponds
to the contribution of low-lying resonance which is not described
by PT it seems unjustified to neglect the contributions from 
high-dimension condensates 
and use only the PT expressions (supplemented by 
condensates only up to dimension 6 which are available) for 
large $k$ moments. 
In the analysis presented in ref.~\cite{pichprades} the 
contributions from high-dimension condensates (with $D>6$)
were regarded as an additional theoretical uncertainty.
Note, however, that the high-dimension condensates 
contribute quite differently to moments $(k,0)$ and $(0,l)$
because the integrals corresponding to these moments 
are saturated by contributions of different energy regions.
The $(0,l)$ moments for large $l$ are saturated at large
energies (of order $M_\tau$) and nonPT contributions
described by the high-dimension condensates within OPE are
small. In the leading order only one condensate is picked up
by integration for a $(0,l)$ moment
(assuming that the factor $(1-s/M_\tau^2)^2$
is removed from the rate).
On the contrary, the $(k,0)$ moments for large $k$ are
saturated at small energies which correspond to the region of
strong coupling: these moments are dominated by the resonances.
Therefore the $(k,0)$ moments are definitely nonperturbative
for large $k$. Within OPE paradigm this means that 
the total contribution of high-dimension condensates is large
compared to perturbative contribution.
Indeed, all condensates with dimension up to $D=2k+8$ contribute to
$(k,0)$ moments and
arrange themselves in a way to reproduce contributions  
of the low-lying resonances according to the standard phenomenology
within OPE. This situation is clearly seen in
exactly solvable models where all power corrections are known.
In QCD, however, numerical values of 
high-dimension condensates are not known and the total contribution
of high-dimension condensates cannot be quantitatively
analyzed for large $k$ moments
though qualitative arguments are rather transparent. 
These reasons forced us to use only the $(0,0)$ moment as the
most reliable from the theoretical
point of view even despite the fact that the experimental
precision for high $k$ moments is better.

\section{Conclusion} 
We have considered the $m_s^2$-corrections 
in a QCD-based description of the $\tau$-system. 
We use a natural effective scheme 
well suited for a contour improved perturbation theory analysis.
The quality of our results is determined by the 
PT expansions of the effective $\beta$- and $\gamma$-functions which
are the only perturbative objects in our analysis.
The $\gamma_q$-function already shows an asymptotic growth 
at NNLO while the $\beta$- and $\gamma_g$-functions 
still 'converge' up to this order.
In our discussion of the N$^3$LO 
terms of PT series for these functions 
(which depend on the unknown parameter $k_3$) we found strong 
indications for asymptotic growth at this order. 
This shows that the ultimate theoretical
limit of FOPT precision is already reached
for the set of observables in the $\tau$-system.
This is not yet an actual problem of QCD because of the 
insufficient precision of the experimental 
data on Cabibbo-suppressed $\tau$ decays especially 
regarding the differential $\tau$ decay rate.
The experimental situation may, however, improve soon.
Then our procedure of using the 
effective scheme description of the $\tau$-system
opens the possibility 
of high precision tests of QCD independently of an explicit 
convergence of PT series in the $\MSsch$ scheme.
In this field an effective scheme approach 
can show its real power
because then the main source of theoretical uncertainty, the relation 
of the internal mass parameters 
$m_q$, $m_g$ to the $\MSsch$ parameter $m_s$, 
will be eliminated. Note that for a QCD test 
within our approach it is necessary to 
relate four $\tau$-observables to each other
because the three parameters
$a$, $m_g$ and $m_q$ have to be fixed.

Our result for the strange quark mass is 
$m_s(\mts)= 130 \pm 27_{\rm{exp}} \pm 9_{\rm th}~{\rm MeV}$
where we have combined the pure theoretical error which is basically 
determined by the truncation of PT series for effective
RG functions and the error
due to the strange quark condensate into one number linearly.
After running to the scale $1~{\rm GeV}$ we obtain
$m_s(1~\rm{GeV})=176 \pm 37_{\rm{exp}}\pm 13_{\rm{th}}~\rm{MeV}$.
This is consistent with the previous results 
where the resummation was done in the $\MSsch$
scheme \cite{msNPB}. The large part 
of the theoretical uncertainty of our result
for the PT coefficient of $m_s^2$ term
comes from re-expressing
the effective quantities in terms of the $\MSsch$ scheme
parameters. 
The advantage of the procedure presented here is that the 
estimate of the accuracy is not based on the 
decomposition of the result into terms coming from 
corrections to the correlator in the $\MSsch$ scheme.
This decomposition seems unnatural 
in resummed perturbation theory as it introduces 
an additional uncertainty related to the convergence of the 
series for the correlator in the $\MSsch$ scheme
every term of which is given by a closed expression resulting from the
resummation along the contour.
Within the effective scheme approach all sources of uncertainty  
are collected into the effective $\beta$- and $\gamma$-functions
which are the only PT quantities entering into the analysis
that provides a solid ground for estimating 
the accuracy of theoretical expressions.

\subsection*{Acknowledgments}
A.A.~Pivovarov thanks K.G.~Chetyrkin for discussion and correspondence.
The work is partially supported by the Russian Fund for Basic Research 
under contracts 99-01-00091 and 01-02-16171 and by the Volkswagen 
Foundation under contract No.~I/73611.

\section*{Appendix}
We give the expressions for the effective $\beta$- and 
$\gamma$-function coefficients in terms 
of the standard $\MSsch$ scheme results. In this Appendix 
the coefficients of $\beta$- and $\gamma$-functions without 
upper index stand for $\MSsch$ quantities while 
the coefficients with upper index ``eff'' denote the 
coefficients of the effective functions.

The coefficients of the effective $\beta$-function
(\ref{betaf}) are given by 
\ba
\beta_0^{\rm{eff}} &=& \beta_0, \qquad
\beta_1^{\rm{eff}} = \beta_1, \nn \\
\beta_2^{\rm{eff}} &=& \beta_2 - k_1 \beta_1 
 + (k_2 - k_1^2)\beta_0, \nn \\
\beta_3^{\rm{eff}} &=& \beta_3 - 2 k_1 \beta_2 + k_1^2 \beta_1
 + (2 k_3 -  6 k_2 k_1 +  4 k_1^3 ) \beta_0 \, .
\ea
For the effective $\gamma$-function coefficients
of eqs.~(\ref{gammagf},\ref{gammaqf}) in terms of 
the $\MSsch$ scheme $\beta$- and 
$\gamma$-function coefficients we find
\ba
\gamma_{n0}^{\rm{eff}} &=& \gamma_0, \qquad
\gamma_{n1}^{\rm{eff}} = \gamma_1 - k_1 \gamma_0 
+ \frac{1}{2} k_{n0}\beta_0, \nn \\
\gamma_{n2}^{\rm{eff}} &=& \gamma_2 - 2 k_1 \gamma_1 
+( - k_2  + 2 k_1^2 ) \gamma_0+\frac{1}{2} k_{n0} \beta_1
+ \left(- k_1 k_{n0}+ k_{n1}- \frac{1}{2} k_{n0}^2 
\right)\beta_0, \nn \\
\gamma_{n3}^{\rm{eff}} &=&  \gamma_3  - 3 k_1 \gamma_2
 +( - 2 k_2 + 5 k_1^2) \gamma_1 
+( - k_3 +  5 k_2 k_1 - 5 k_1^3) \gamma_0 \nn \\
 && +\frac{1}{2} k_{n0} \beta_2 
+ \left( -\frac{3}{2}k_1 k_{n0} +  k_{n1}  -\frac{1}{2} k_{n0}^2 
  \right) \beta_1  \\
 &&  + \left( - k_2 k_{n0}+ \frac{5}{2} k_1^2 k_{n0}
 - 3 k_1 k_{n1} + \frac{3}{2} k_1 k_{n0}^2 + \frac{3}{2} k_{n2} 
-\frac{3}{2}k_{n1} k_{n0} +\frac{1}{2} k_{n0}^3 \right)\beta_0 .\nn
\ea
Here $k_{nj}$ with $n = q,g$ stand for the coefficients
of eq.~(\ref{effmg}) if $n=g$ $(k_{g0} = 5/3)$ and for the coefficients
of eq.~(\ref{effmq}) if $n=q$ $(k_{q0} = 7/3)$.
Numerical values for all necessary
coefficients in the $\MSsch$ scheme have been collected in 
\cite{onetwo} where further references to original papers 
can be found.

\end{document}